\documentclass[amssymb, nobibnotes, aps, prd]{revtex4-1}
\usepackage{epsfig}
\usepackage{amsmath}
\usepackage{amsfonts}
\usepackage{amssymb}
\usepackage{graphicx}
\usepackage{colordvi}

\usepackage{amsthm}
\usepackage{slashed}
\usepackage{bbm}
\usepackage{color}
\usepackage{amsfonts}

\usepackage{array}

\begin{document}
\begin{flushleft}
KCL-PH-TH/2016-03
\end{flushleft}

\title{Linear stability of noncommutative spectral geometry}

\author{M.\ Sakellariadou\footnote{email address:
    mairi.sakellariadou@kcl.ac.uk} and
  A.\ Watcharangkool\footnote{email address:
    apimook.watcharangkool@kcl.ac.uk}}

\affiliation{Department of Physics, King's College London, University
  of London, Strand WC2R 2LS, London, United Kingdom}

\begin{abstract}
We consider the spectral action within the context of a 4-dimensional
manifold with torsion and show that, in the vacuum case, the equations
of motion reduce to Einstein's equations, securing the linear stability
of the theory.  To subsequently investigate the nonvacuum case, we
consider the spectral action of an almost commutative torsion geometry
and show that the Hamiltonian is bounded from below, a result which
guarantees the linear stability of the theory.
\end{abstract}
\maketitle

\section{Introduction}
Consider the gravitational action
\begin{equation}
 S_{\rm gr}[g_{\mu\nu}] = \int{ \sqrt{|g|}}\left(\bar{\Lambda}
 +\frac{1}{\kappa^2} R-\alpha_0 ||C||^2\right)d^4x~,\label{eq:Higher}
\end{equation}
where $\bar{\Lambda}$ denotes the cosmological constant,
$\kappa^2=16\pi G$, $\alpha_0$ is a positive constant and
$||C||^2:=C^{\mu\nu\rho\sigma}C_{\mu\nu\rho\sigma}$ is the Weyl
invariant. This action, Eq.~(\ref{eq:Higher}), belongs to a family of
higher derivative theories, since it contains a fourth order
derivative of the metric tensor, $g_{\mu\nu}$, namely $(\partial^2
g)^2$. The presence of this higher derivative term may give rise to an
unbounded (from below) Hamiltonian, implying the onset of a classical
instability~\cite{Stelle, Hans, Eugene}. Gravitational theories with
curvature invariants, as for instance shown in the
action~(\ref{eq:Higher}) above, belong to the class of nondegenerate
higher derivative theories plagued by the Ostrogradski instability
(linear instability).  Such theories  can appear naturally in the context of fundamental
theories, as for instance within one-loop corrections of quantum
theories on a curved background, or within the spectral action of
almost commutative geometry. Despite the fact that  theories with higher
derivative terms may be pathological, they may instead improve the ultraviolet convergence of the
graviton propagator within a linearized theory, rendering the theory
power counting renormalizable~\cite{Stelle}.

The linear instability of a higher derivative nondegenerate theory may
be removed, if one assumes the action as an effective one~\cite{Simon}
and imposes appropriate constraints leading to a reduction of the
trajectories of the degrees of freedom, hence rendering the effective
Hamiltonian bounded from below. In this approach, a necessary but not
sufficient condition in order to remove the instability, is that the
imposed constraints must be such that they reduce the dimensionality
of the original phase space~\cite{Eugene}. A different approach has
been suggested in Ref.~\cite{Wheeler}, where one generalizes a higher
derivative theory into an SO(4,2) gauge theory, and then derives
conditions such that the equations of motion reduce, in some basis, to
the vacuum Einstein's equations. As it has been shown~\cite{Wheeler},
varying all of the connection fields, and not only the metric, Weyl
gravity transforms from a fourth order theory into a theory of
conformal equivalence classes of solutions to general relativity,
under the requirement that torsion vanishes.  In what follows, we show
the linear stability of the spectral action for a 4-dimensional
manifold with torsion and in the absence of any matter fields,
adapting the approach proposed in Ref.~\cite{Wheeler}.  We
subsequently extend this approach in the nonvacuum case.

This paper is organized as follows. In Section~2, we briefly
introduce~\cite{ncg-book1,ncg-book2,Walter,Sakellariadou:2012jz} the
concept of spectral geometry and spectral action. In Section~3, we
review the approach discussed in Ref.~\cite{Wheeler}, and apply it to
the fourth order gravitational theory described by the action
(\ref{eq:Higher}). We show that such theory does not suffer from
linear instability. In Section~4, we consider the spectral action of
an almost commutative torsion geometry and show that the obtained
Hamiltonian is bounded from below, hence the theory does not suffer
from linear instability even in the nonvacuum case. We round up our
conclusions in Section~5.
\section{Elements of the spectral action}

Consider a compact 4-dimensional Riemannian spin manifold $M$ and a
spinor bundle $S\rightarrow M$. The set of smooth, infinitely
differentiable, functions $C^\infty(M)$ forms an algebra $A$ under
pointwise multiplication.  This algebra acts on the Hilbert space of
square-integrable spinors on $M$, $H=L^2(M,S)$, as multiplication
operators.  Then consider the Dirac operator $\mathcal{D}$, given in
terms of the spin Levi-Civita connection $\nabla^S$ and the Dirac
gamma matrices $\gamma^\mu$ as $-i\gamma^\mu\nabla^S_\mu$.  The
compact Riemannian spin manifold $M$ is fully described~\cite{C-Man}
by the canonical spectral triple $(A, H,\mathcal{D})$. Hence, spectral
data can characterize the geometry of ordinary Riemannian manifolds,
in the sense that the canonical spectral triple $(A, H,\mathcal{D})$
encodes the spacetime structure.

Let us extend the spectral triple approach for noncommutative manifolds.
Conside the finite $C^*$-algebra
\begin{equation}
A_F=\mathbbm{C}\oplus\mathbbm{H}\oplus M_4(\mathbbm{C}),
\end{equation}
together with a finite dimensional Hilbert space $H_F$ and a
self-adjoint operator $D_F$. The spectral triple $(A_F,H_F,D_F)$ can be
identified with a finite space of points $F$. Although the finite
spectral triple by itself gives an uninteresting structure, its product
with the canonical spectral triple, namely
\begin{equation}
(C^\infty(M)\otimes A_F,L^2(M,S)\otimes H_F, \slashed{\nabla}\otimes
  \textup{Id}_F+\gamma^5\otimes D_F), \label{eq:AC-man}
\end{equation} 
yields a nontrivial noncommutative structure~\cite{Walter}. The
spectral triple (\ref{eq:AC-man}) is called the almost commutative
spectral triple. The canonical triple encodes the spacetime structure
while the finite spectral triple encodes the internal degrees of
freedom at each point of the 4-dimensional spacetime.  The particle
physics model one has in mind is encoded in the finite dimensional
Hilbert space $H_F$. In the case of the Standard Model, the
generalized Dirac operator acting on the Hilbert space
$H=L^2(M,S)\otimes H_F$, contains the Higgs boson, Yukawa couplings,
neutrino masses, as well as the Cabibbo-Kobayashi-Maskawa matrix.

The dynamics are given by a spectral action that sums up all
frequencies of vibration of space. The spectral action is defined as
the heat kernel trace of the operator $\mathcal{D}^2$:
 \begin{equation}
 S= \textup{Tr}_{L^2} f(\mathcal{D}^2/\Lambda^2)~,
 \end{equation} 
 where $f$ is a positive cut-off function and $\Lambda$ a cut-off
 scale. For the canonical spectral triple the spectral action reads
\begin{align}
S \sim &\int{\sqrt{|g|}}\left(\frac{f_4}{2\pi^2}\Lambda^4 +
\frac{f_2}{24\pi^2}\Lambda^2 R-\frac{f(0)}{16\pi^2}
||C||^2\right) d^4x+{\cal O}(\Lambda^{-2}), \label{eq:SA-can}
\end{align}
where 
\begin{equation}
f_{4-k}=\int^\infty_0 x^{4-k-1}f(x)dx~~,~~0\leq k<4~.
\end{equation} 
Note that the action (\ref{eq:SA-can}) is of the same form as the
action (\ref{eq:Higher}), which is a higher derivative gravitational
theory.  In the case of the almost commutative spectral triple, the
spectral action reads~\cite{Walter}
\begin{align}
S \sim &\int{\sqrt{|g|}}\left[\frac{48 f_4}{\pi^2}\Lambda^4-\frac{c
    f_2}{\pi^2}\Lambda^2+\frac{d
    f(0)}{4\pi^2}\frac{4f_2}{\pi^2}\Lambda^2-\frac{cf(0)}{24\pi^2}
  R-\frac{3f(0)}{10\pi^2} ||C||^2\right.\nonumber\\ & \left.
  \ \ \ \ \ \ \ \ \ \ +\frac{1}{4}F_{\mu\nu}F^{\mu\nu} +
  \frac{1}{4}W_{\mu\nu}^a W^{\mu\nu, a} + G_{\mu\nu}^i G^{\mu\nu,
    i}\frac{1}{2}|\nabla'_\mu H|^2-\frac{1}{12}RH^2
  \right.\nonumber\\ &
  \left. \ \ \ \ \ \ \ \ \ \ -\frac{2af_2\Lambda^2-ef(0)}{af(0)}H^2
  +\frac{b\pi^2}{2a^2f(0)} H^4 \right]d^4x+{\cal O}(\Lambda^{-2})~,
\end{align}           
where the action of $~\nabla'$ on the Higgs field is defined as 
\begin{equation}
\nabla'_\mu H := \partial_\nu H+ \frac{1}{2}ig_2 W_\mu^a\sigma^a
H-\frac{1}{2}ig_1 A_\mu H~.
\end{equation}
and the constants $a,b,c,d$ and $e$ are derived from Yukawa mass
matrices. The gauge fields $A_\mu, W_\mu$ and $G_\mu$ belong to the
Lie algebra of the symmetry groups U$_Y(1)$, SU(2) and SU(3),
respectively.
\section{Fourth order Weyl gravity}
Consider the higher derivative theory 
\begin{equation}
S=\int{\mathbf{\Omega}^A_{\ \ B}}\wedge ^ *\mathbf{\Omega}^B_{\ \ A}~,
\label{act-4ord}
\end{equation}
where $\mathbf{\Omega}^A_{~B}$ stands for the SO(4,2) curvature
2-form.  As it has been shown in Ref.~\cite{Wheeler}, varying the
above action with respect to the connection, the higher order equations of
motion can be reduced, in the absence of torsion,  to the
vacuum second order Einstein's equations. The solutions are conformal
equivalence metrics of Ricci-flat spacetimes. Following this approach
for the generalized spectral action which is invariant under a smaller
symmetry group, i.e. local Poincar$\acute{\text{e}}$ symmetry, we will
show that in the absence of torsion the equations of motion combined
with the Bianchi identity lead to an integrability condition that
implies the reduction to the second order Einstein's equations.

To generalize the action Eq.~(\ref{eq:SA-can}) into a gauge theory
with a Poincar$\acute{\text{e}}$ symmetry one needs to equip a
manifold with a tetrad $e^a_\mu$,
\begin{equation}
 g_{\mu\nu}=\eta_{ab}e^a_\mu e^b_\nu~, 
 \end{equation}
and a spin connection $\omega^{ab}_\mu \in \mathfrak{so}(1,3)$, satisfying
\begin{equation}
D_\mu e_\nu^a:=\partial_\mu
e^a_\nu-\Gamma^\alpha_{\mu\nu}e^a_\alpha+\omega^a_{\mu~c}
e^c_\nu=0~,\label{eq:tetrad}
 \end{equation}
where latin characters denoting flat spacetime indices, $D_\mu$ is the
covariant derivative and $\Gamma^\alpha_{\mu\nu}$ is an affine
connection. The curvature two-form of the spin connection, defined by
 \begin{equation}
 R_{\mu\nu}^{\:\:\:\: ab}:=
 \partial_{\mu}\omega^{ab}_{\nu}-\partial_{\nu}\omega^{ab}_{\mu}+\omega^a_{\mu
   c}\omega^{cb}_\nu - \omega^a_{\nu
   c}\omega^{cb}_\mu~,\label{eq:curvature}
 \end{equation}
is independent of the tetrad basis. In general, the spin connection is not
necessarily torsion-free. In fact, the curvature two-form
(\ref{eq:curvature}) contains a torsion and its derivative.  This can
be shown by contracting Eq.~(\ref{eq:tetrad}) with $e^{\nu,b}$,
\begin{align}
\omega_\mu^{ab}=&~e^a_\nu
e^{\sigma,b}\Gamma^\nu_{\mu\sigma}+e^a_\nu\partial_\mu e^{\nu,
  b}\nonumber\\ =&~e^a_\nu
e^{\sigma,b}\left(\Gamma^\nu_{(\mu\sigma)}+\Gamma^\nu_{[\mu\sigma]}\right)
+e^a_\nu\partial_\mu
e^{\nu, b}\nonumber\\ =& \left(e^a_\nu
e^{\sigma,b}\Gamma^\nu_{(\mu\sigma)}+e^a_\nu\partial_\mu e^{\nu,b}
\right)+e^a_\nu e^{\sigma,b}\Gamma^\nu_{[\mu\sigma]}\nonumber\\ =&
\left(e^a_\nu e^{\sigma,b}\Gamma^\nu_{(\mu\sigma)}+e^a_\nu\partial_\mu
e^{\nu,b} \right)+\frac{1}{2}e^a_\nu
e^{\sigma,b}T^{~\nu}_{\mu~\sigma}~, \label{eq:rmk1}
\end{align}
where $T^{~\nu}_{\mu~\sigma}:=2\Gamma^\nu_{[\mu\sigma]}$ is the
torsion tensor. The subscript notation ``$(~~)$" denotes
symmetrization
$\Gamma^\nu_{(\mu\sigma)}:=\frac{1}{2}(\Gamma^\nu_{\mu\sigma}
+\Gamma^\nu_{\sigma\mu})$
and ``$[~~]$" donotes anti-symmetrization
$\Gamma^\nu_{[\mu\sigma]}:=\frac{1}{2}(\Gamma^\nu_{\mu\sigma}
-\Gamma^\nu_{\sigma\mu})$.

Defining 
\begin{equation}
\omega_\mu'^{~ab}:= e^a_\nu
e^{\sigma,b}\Gamma^\nu_{(\mu\sigma)}+e^a_\nu\partial_\mu e^{\nu,b}~,
\end{equation}
we note that $\omega'$ is torsion-free and the curvature
(\ref{eq:curvature}) can be rewritten as
\begin{align}
R_{\mu\nu}^{\:\:\:\:ab}=& R'^{~~ab}_{\mu\nu}+\nabla_\mu
T^{~ab}_\nu-\nabla_\nu T^{~ab}_\mu+T_{\mu~c}^{~a} T^{~cb}_\nu
-T_{\nu~c}^{~a} T^{~cb}_\mu ~, \label{eq:Riem-T}
\end{align}
where $\nabla$ is a covariant derivative acting on a tensor
  $v^{~a}_\nu$ as
\begin{equation}
\nabla_\mu v^{~a}_\nu:=\partial_\mu
v^a_\nu-\Gamma^\alpha_{(\mu\nu)}v^a_\alpha+\omega'^a_{\mu~c}v^c_\nu~,
\end{equation} 
and $R_{\mu\nu}'^{~~ab}$ is the curvature two-form of the torsion-free
spin connection $\omega'^{ab}_{\mu}$, defined by
\begin{equation}
R_{\mu\nu}'^{~~ab}:=
\partial_{\mu}\omega'^{ab}_{\nu}-\partial_{\nu}\omega'^{ab}_{\mu}+\omega'^a_{\mu
  c}\omega'^{cb}_\nu - \omega'^a_{\nu c}\omega'^{cb}_\mu~.
\end{equation} 
Denoting by $\mathcal{T}$ the set of all torsion fields, we consider a
particular subset $\mathcal{T}_R \subset \mathcal{T}$, so that the
torsion fields $~T_\mu^{~ab} \in \mathcal{T}_R~$ satisfy the following
properties:\\ $\bullet$ $~T_\mu^{~ab}$ is antisymmetric in the $a, b$
indices, and hence Eq.~(\ref{eq:rmk1}) implies that
$\omega_\mu'^{~ab}$ is also antisymmetric in $a, b$, leading to metric
compatibility, and $\omega_\mu'^{~ab}$ is just the Levi-Civita
connection.  The reason for choosing totally antisymmetric torsion
fields is the following: The general connection on the tangent bundle
of a manifold is compatible with the Riemannian metric and has the
same geodesics as the Levi-Cevita connection if and only if the
connection is the sum of the Levi-Cevita connection and a totally
antisymmetric tensor field~\cite{TSA}, thus the torsion field is totally
antisymmetric.  \\ $\bullet$ $T_\mu^{~ab}$ yields the curvature tensor
with the same symmetric properties as the Riemmanian curvature tensor,
i.e.
\begin{align}
R_{\mu\nu\sigma\rho}=-&R_{\nu\mu\sigma\rho}=R_{\nu\mu\rho\sigma}~, \label{eq:sym-1}
\\ R_{\mu\nu\sigma\rho}& = R_{\sigma\rho\mu\nu} \label{eq:sym-2}~,
\end{align}
where $R_{\mu\nu\sigma\rho}=R_{\mu\nu}^{\ \ \ ab} ~e_{\sigma,
  a}~e_{\rho, b}$. Note that (\ref{eq:sym-1}) holds for all torsion
fields $T_\mu^{~ab} \in \mathcal{T}$, while (\ref{eq:sym-2}) is only
valid for $T_\mu^{~ab} \in \mathcal{T}_R$.  With the above properties
of the torsion fields, the Gauss-Bonnet action takes the form we are
familiar with in Riemannian geometry, namely
\begin{equation}
\chi_E=\frac{1}{8\pi^2}\int{\sqrt{|g|}}(R_{\mu\nu}^{~~ab}
R^{\mu\nu}_{~~ab}-4R_\mu^{~a}R^\mu_{~a}+R^2) d^4x~.\label{eq:GB-T}
\end{equation}
We note that the above action~(\ref{eq:GB-T}) is not valid for the more general class
of torsions studied in Ref.~\cite{gr-TSA}.

Let us also define a traceless tensor $C_{\mu\nu}^{\:\:\:\:ab}$, as
\begin{equation}
C_{\mu\nu}^{\:\:\:\:ab}:=~R_{\mu\nu}^{\:\:\:\:ab}-(e^{[a}_\mu
  R^{~b]}_\nu-e^{[a}_\nu R^{~b]}_\mu)+\frac{1}{3}Re^{[a}_\mu
  e^{b]}_\nu~,\label{eq:Weyl-Def}
\end{equation}
where $R^{~a}_\mu:=~R_{\mu\nu}^{\:\:\:\:ab}e^\nu_b$ and $R:=R^{~a}_\mu
e^a_\mu$. We can thus generalize the spectral action
Eq.~(\ref{eq:SA-can}) as follows:
\begin{align}
 S_{\rm gr}[e^a_\mu, \omega^{ab}_\nu] = \int e\left(\alpha_2\Lambda
 ^4+\frac{1}{\kappa^2} R_{\mu\nu}^{\:\:\:\:ab}e_a^\mu e_b^\nu-\alpha_0
 C_{\mu\nu}^{\:\:\:\:ab}C^{\mu\nu}_{\:\:\:\:ab}\right)d^4x~, \label{eq:NCG
   Kibble}
 \end{align}
 where $e$ is defined as $e:=\sqrt{|\det(e_\mu^a
   e_{a,\nu})|}=\sqrt{|g|}$.  For a torsion field $~T_\mu^{~ab}\in
 \mathcal{T}_R~$, it can be shown that the linearized theory obtained
 from the action (\ref{eq:NCG Kibble}) is equivalent to the one
 derived by the spectral action with torsion~\cite{gr-TSA} (see
 Appendix~B). Hence, the action (\ref{eq:NCG Kibble}) is linearly
 stable if and only if the spectral action with torsion is linearly
 stable. 

Let us now derive the equations of motion obtained from the
   generalized action (\ref{eq:NCG Kibble}). The variation of the
   spin connection and the tetrad give respectively,
 \begin{align}
 D_\mu C^{\mu\nu}_{\:\:\:\:ab}-\frac{1}{2}T^{~\nu}_{\mu~\alpha}
 C^{\mu\alpha}_{\:\:\:\:ab}=&~-\frac{1}{4\alpha_0\kappa^2}T^{~\nu}_{\mu~\alpha}e^\mu_a
 e^\alpha_b~,
 \\ R_{\mu\nu}-\frac{1}{2}g_{\mu\nu}R=&~2\alpha_0\kappa^2\Theta_{\mu\nu}
 + \frac{\kappa^2}{2}\alpha_2\Lambda^4g_{\mu\nu}~,\label{eq:Scalar-T}
 \end{align}
where $\Theta_{\mu\nu}:=
C_{\mu\alpha}^{\:\:\:\:ab}C_{\nu\:\:\:ab}^{\:\:\alpha}
-\frac{1}{4}g_{\mu\nu}C_{\mu\nu}^{\:\:\:\:ab}C^{\mu\nu}_{\:\:\:\:ab}$. To
recover Einstein's equations from Eq.~(\ref{eq:Scalar-T}) we need
first to set the torsion equal to zero, so that the connection becomes
the Levi-Civita one. Thus,
 \begin{align}
 \nabla_\mu
 C'^{\mu\nu}_{\:\:\:\:ab}=&~0~, \label{eq:Weyl}\\ 
R'_{\mu\nu}-\frac{1}{2}g_{\mu\nu}R'=&~2\alpha_0\kappa^2\Theta'_{\mu\nu}
 + \frac{\kappa^2}{2}\alpha_2\Lambda^4g_{\mu\nu}~, \label{eq:Scalar}
 \end{align}
where $\Theta'_{\mu\nu}:=\Theta_{\mu\nu}|_{T=0}$. Since
$\Theta'_{\mu\nu}$ becomes the energy momentum tensor of the Weyl
curvature, and therefore vanishes identically in
4-dimensions~\cite{Juan}, we recover Einstein's equations with a
cosmological constant.

The vanishing divergence of the Weyl curvature, Eq.~(\ref{eq:Weyl}),
leads to the integrability condition once combined with the trace of
the Bianchi identity
 \begin{equation}
 \nabla_\mu C'^\mu_{\:\:\nu\rho\sigma}+(\nabla_\sigma
 S'_{\nu\rho}-\nabla_\rho S'_{\nu\sigma})=0~,\label{eq:Bianchi}
 \end{equation}
 where
 $S'_{\mu\nu}:=\frac{1}{2}\left(R'_{\mu\nu}-\frac{1}{6}g_{\mu\nu}R'\right)$
 denotes the Schouten tensor. In the
 basis $e^a_\mu$, we get
\begin{equation}
\nabla_\sigma S'_{\nu\rho}-\nabla_\rho S'_{\nu\sigma}=0~,
\end{equation} 
which however is not the well-known integrability condition. To get
the familiar expression~\cite{inte} we introduce a new basis $e^a_\mu
\mapsto \tilde{e}^a_\mu:=e^\xi e^a_\mu$, where $\xi(x)$ is a
real-value function. Note that the Bianchi identity holds in this new
basis, but the covariant derivative of the Weyl tensor transforms as
 \begin{equation}
 \tilde{\nabla}_\mu \tilde{C'}^\mu_{\:\:\nu\rho\sigma}=
 e^{-2\xi}\left(\nabla_\mu C'^\mu_{\:\:\nu\rho\sigma}-\partial_\mu \xi
 C'^\mu_{\:\:\nu\rho\sigma}\right)~.\label{eq:C-trans}
 \end{equation}
To get the integrability condition, we consider Eq.~(\ref{eq:Bianchi})
  in the basis $\tilde{e}^a_\mu$ and use Eq.~(\ref{eq:C-trans}) and 
the field equation (\ref{eq:Weyl}), to obtain
\begin{align}
0=& \tilde{\nabla}_\mu
\tilde{C'}^\mu_{~\nu\rho\sigma}+(\tilde{\nabla}
_\sigma\tilde{S}'_{\nu\rho}-\tilde{\nabla}_\rho
\tilde{S}'_{\nu\sigma})\nonumber\\ =&e^{-2\xi}\left(\nabla_\mu
C'^\mu_{~~\nu\rho\sigma}-\partial_\mu \xi
C'^\mu_{~~\nu\rho\sigma}\right)+(\tilde{\nabla}_\sigma\tilde{S}'_{\nu\rho}
-\tilde{\nabla}_\rho\tilde{S}'_{\nu\sigma})\nonumber\\ =&
-(\partial_\mu \xi
)e^{-2\xi}C'^\mu_{~~\nu\rho\sigma}+(\tilde{\nabla}_\sigma\tilde{S}'_{\nu\rho}
-\tilde{\nabla}_\rho\tilde{S}'_{\nu\sigma})
\nonumber\\ =&-(\partial_\mu \xi)\tilde{C'}_{~\nu\rho\sigma}^{\mu}+
\tilde{\nabla}_{\sigma} \tilde{S'}_{\rho\nu}-\tilde{\nabla}_{\rho}
\tilde{S'}_{\sigma\nu}~,\label{eq:int. cond}
 \end{align}
where we have used that
$e^{-2\xi}C'^\mu_{~~\nu\rho\sigma}=\tilde{C}'^\mu_{~~\nu\rho\sigma}$. Hence,
the original manifold is conformally equivalent to a Ricci
    flat manifold. In other words, there exists a basis $\hat{e}^a_\mu:=e^\zeta
    \tilde{e}^a_\mu$, equal to $\hat{e}^a_\mu=e^{\xi+ \zeta} e^a_\mu$  such that
 \begin{equation}
 \hat{S}'_{\mu\nu}=0~, \label{eq:Rf}
 \end{equation}
 leading to vanishing Ricci tensor, $\hat{R}'_{\mu\nu}=0$. Therefore,
 the equation of motion (\ref{eq:Weyl}) is conformally equivalent to
 the vacuum Einstein's equations and the theory is not plagued by a
 linear instability. Defining $\bar\chi:=\xi+\zeta$, the Schouten tensor reads
 \begin{equation}
\hat{S}'_{\mu\nu}=S'_{\mu\nu}-\nabla_\mu\partial_\nu \bar\chi +
\partial_\mu\bar\chi
\partial_\mu\bar\chi-\frac{1}{2}g_{\mu\nu}\partial^\alpha\bar\chi
\partial_\alpha \bar\chi~,
 \end{equation}
and Eq.~(\ref{eq:Rf}) is compatible with Eq.~(\ref{eq:Scalar})
providing the scalar field $\bar{\chi}$ satisfies
\begin{equation}
\nabla_\mu\partial_\nu \bar\chi - \partial_\mu\bar\chi
\partial_\nu\bar\chi-g_{\mu\nu}\left(\nabla_\alpha\partial^\alpha\bar\chi+\frac{1}{2
}\partial^\alpha\bar\chi \partial_\alpha
\bar\chi\right)=\frac{1}{4}\kappa^2\alpha_2\Lambda^4
g_{\mu\nu}~. \label{eq:Conf.}
\end{equation}
In conclusion, considering the variation of the full connection, the
higher order differential equations reduce to Einstein's equations
obtained from either Eq.~(\ref{eq:Weyl}) or from
Eq.~(\ref{eq:Scalar}).
\section{Hamiltonian analysis of the theory interacting with matter fields}

Let us now assume that the gravitational action is defined
in a 4-dimensional globally hyperbolic manifold, of the
structure $~\mathbbm{R}\times \Sigma~$, where $\Sigma$ is a Cauchy
surface, i.e. any curve parametrized by $t\in \mathbbm{R}$
intersects $\Sigma$ only once~\cite{Cauchy}. Consequently, if one
picks the time direction along a normal vector on a Cauchy surface, 
there is no closed time-like curve in the manifold. More importantly,
the existence of a Cauchy surface at any instant of time allows us to
define the Poisson bracket, which is important for setting the
Hamiltonian formalism.

Global hyperbolicity also allows us to choose a coordinate system $\{t,
x^i\}$ such that the spatial coordinates  are orthogonal to the time
coordinate, i.e. $g_{ti}=0$. Let us choose flat spacetime basis
$\{\mathbf{e}^0, \mathbf{e}^I\}$ with $I\in \{1,2,3\}$, such that the time
direction is preserved
\begin{equation}
\mathbf{e}^0=e^0_t dt  ~~~ \mbox{and} ~~~  \mathbf{e}^I=e^I_i dx^i~.
\end{equation}
In the previous section we have avoided the linear instability by
conformally reducing the equations of motion (\ref{eq:Weyl}) to the
vacuum Einstein's equations. The same method can be extended to the
nonvacuum case as long as $\mathcal{L}_{\rm matter}$ is not a function
of the spin connection, as for instance for the Lagrangian of a gauge
field. Note however that there are matter fields whose Lagrangian
depends on the spin connection, as for example
 \begin{align}
\mathcal{L}_H=&~\frac{1}{2}|\nabla'_\mu
H|^2-\frac{1}{12}RH^2-\mu^2H^2+\lambda H^4~, \\ \mathcal{L}_\psi=&~
i\bar{\psi}(e^\mu_a\gamma^a D_\mu - m)\psi~,
\end{align}
where $\nabla'_\mu H=\partial_\mu H +[B_\mu,H]$ and
$D_\mu\psi:=(\partial_\mu +
\frac{1}{4}\omega_\mu^{ab}\Sigma_{ab})\psi$, for
$\Sigma_{ab}:=\frac{1}{2}(\gamma_a\gamma_b-\gamma_b\gamma_a)$. Such
Lagrangians lead to the equations of motion
 \begin{equation}
 \nabla_\mu C'^{\mu\nu}_{\:\:\:\:ab}=\frac{\delta \mathcal{L}_{\rm
     matter}}{\delta \omega^{ab}_\nu}\bigg|_{T=0}\not=0~.
 \end{equation}
In such a case one cannot get the integrability condition using the
same trick as previously, and hence one cannot argue the cure of the
linear instability following the approach of Section~2. To show that
there is no instability we will check directly that the Hamiltonian
is bounded from below.

Without loss of generality let us turn off the gauge fields and the
cosmological constant since they do not depend on the spin
connection. By adding the Higgs field and a massive fermionic field
into the action (\ref{eq:NCG Kibble}) we get
\begin{align}
 S_{\rm gr}[e^a_\mu, \omega^{ab}_\nu]+S_{H,\psi} = \int{d^4x
   e}\left(\frac{1}{\kappa^2} R_{\mu\nu}^{\:\:\:\:ab}e_a^\mu
 e_b^\nu-\alpha_0
 C_{\mu\nu}^{\:\:\:\:ab}C^{\mu\nu}_{\:\:\:\:ab}+\mathcal{L}_{H,\psi}\right)~.
\label{eq:prac}
 \end{align}
The canonical momenta are
\begin{align}
\pi^{\beta}_{cd}&=-4\alpha_0C^{t\beta}_{\:\:\:\:cd}+\left(\frac{2}{\kappa^2}
-\frac{H^2}{6}\right)e^t_{[c}e^\beta_{d]}~,
\\ p^t_0&=0~, ~p^i_I=0~,
\end{align}
where $\pi^\beta_{cd},\ p^t_0$ and $p^i_I$ stand for the canonical
momenta of $\omega^{\:cd}_\beta,\ e^0_t$ and $e^I_i$,
respectively. Notice that the map $\pi_\beta^{ab} \mapsto \partial_t
\omega_\beta^{ab}$ is not invertible for an arbitrary choice of the
spin connection, therefore the Hamiltonian is not well-defined. To
construct a well-defined Hamiltonian, let us consider a subset of spin
connections such that each element can be decomposed into
$\omega_\mu^{\:ab}=\Omega_\mu^{\:ab}+\tilde{\omega}_\mu^{\:ab}$ and
the following two conditions are satisfied:
\begin{flalign}
i)&~~C_{\mu\nu}^{\:\:\:\:ab}=\partial_\mu
\Omega_\nu^{\:ab}-\partial_\nu\Omega_\mu^{\:ab}+\Omega_{\mu\:c}^{\:a}
  \Omega_\nu^{\:cb}-\Omega_{\nu\:c}^{\:a}
  \Omega_\mu^{\:cb}~,\label{eq:sp-1}\\
  ii)&~~ (\Omega_{[\mu}^{\:ac}
  \tilde{\omega}_{\nu] c}^{~b}+\tilde{\omega}_{[\mu}^{~ac}
  {\Omega}_{\nu]c}^{~b})e^\mu_b e^\nu_a=0~. \label{eq:sp-2}&&
\end{flalign}
We will call $i)$ and $ii)$ the ``splitting conditions" since they
make the scalar curvature independent of $\Omega^{~ab}_\mu$. To see
this we rewrite the curvature $R=R_{\mu\nu}^{~~ab}e^\mu_a e^\nu_b$ in
terms of $\Omega$ and $\tilde{\omega}$. Thus,
\begin{align}
R_{\mu\nu}^{\:\:\:\:ab}=&~\partial_\mu \Omega_\nu^{\:ab}-\partial_\nu
\Omega_\mu^{\:ab}+\Omega_{\mu\:\:c}^{\:a}\Omega^{\:cb}_\nu
-\Omega_{\nu\:\:c}^{\:a}\Omega^{\:cb}_\mu
\nonumber\\ &+\partial_\mu\tilde{\omega}_\nu^{\:ab}
-\partial_\nu\tilde{\omega}_\mu^{\:ab}
+\tilde{\omega}_{\mu\:\:c}^{\:a}\tilde{\omega}^{\:cb}_\nu
-\tilde{\omega}_{\nu\:\:c}^{\:a}\tilde{\omega}^{\:cb}_\mu
\nonumber\\ &-2(\Omega_{[\mu}^{\:ac} \tilde{\omega}_{\nu]
  c}^{~b}+\tilde{\omega}_{[\mu}^{~ac} {\Omega}_{\nu]c}^{~b})~.
\end{align} 
Assuming the validity of the conditions $i)$ and $ii)$ above, the
scalar curvature reads
\begin{align}
R=&~R_{\mu\nu}^{~~ab}e^\mu_a
e^\nu_b\nonumber\\ =&~C_{\mu\nu}^{~~ab}e^\mu_a e^\nu_b+
(\partial_\mu\tilde{\omega}_\nu^{\:ab}-\partial_\nu\tilde{\omega}_\mu^{\:ab}
+\tilde{\omega}_{\mu\:\:c}^{\:a}\tilde{\omega}^{\:cb}_\nu
-\tilde{\omega}_{\nu\:\:c}^{\:a}\tilde{\omega}^{\:cb}_\mu)e^\mu_a
e^\nu_b \nonumber \\ &-2(\Omega_{[\mu}^{\:ac} \tilde{\omega}_{\nu]
  c}^{~b}+\tilde{\omega}_{[\mu}^{~ac} {\Omega}_{\nu]c}^{~b})e^\mu_a
e^\nu_b
\nonumber\\ =&~(\partial_\mu\tilde{\omega}_\nu^{\:ab}
-\partial_\nu\tilde{\omega}_\mu^{\:ab}
+\tilde{\omega}_{\mu\:\:c}^{\:a}\tilde{\omega}^{\:cb}_\nu
-\tilde{\omega}_{\nu\:\:c}^{\:a}\tilde{\omega}^{\:cb}_\mu)e^\mu_a
e^\nu_b ~.
\end{align}
Note that the considered subset of spin connections is not empty,
since it contains connections of all conformal Ricci flat
geometry. Moreover, the splitting conditions hold automatically in the
linearized theory. To prove this statement, let $h_{\mu\nu}$ denote
the metric perturbation. The condition $(ii)$ is clearly satisfied
since $\Omega_{\mu~c}^{~a}\tilde{\omega}_\nu^{~cb} e^\mu_a e^\nu_b$ is
of order higher than $O(h^2)$. For condition $(i)$ one chooses the
transverse traceless metric perturbation $\bar{h}_{\mu\nu}$
which satisfies the Laplace equation
\begin{equation}
\Box \bar{h}_{\mu\nu}=0~,
\end{equation}
where $\Box$ denotes the flat space d'Alembertian. The Weyl tensor is
\begin{align}
C_{\mu\nu\sigma\rho}=&\frac{1}{2}(\partial_\sigma \partial_\nu
\bar{h}_{\mu\rho} +\partial_\rho \partial_\mu \bar{h}_{\nu\sigma}
-\partial_\rho \partial_\nu \bar{h}_{\mu\sigma}-\partial_\sigma
\partial_\mu \bar{h}_{\nu\rho})
\nonumber\\ =&\eta_{\mu\lambda}\partial_\sigma
\bar{\Gamma}^\lambda_{\nu\rho}-\eta_{\mu\lambda}\partial_\rho
\bar{\Gamma}^\lambda_{\nu\sigma}~,
\end{align}
where $~\bar{\Gamma}^\lambda_{\nu\rho}:=\frac{1}{2}\eta^{\lambda\mu}(
\partial_\nu \bar{h}_{\rho\mu}+\partial_\rho
\bar{h}_{\nu\mu}-\partial_\mu \bar{h}_{\nu\rho})$. Then using the definition of the spin connection, one can rewrite the Weyl tensor in terms of derivative of $\Omega_\mu^{~ab}$, and therefore the condition $(i)$ is satisfied.

Defining
\begin{equation}
\Pi^\beta_{cd}:=\frac{\partial \mathcal{L}}
{\partial(\partial_t\Omega^{cd}_\beta)}~~~ \mbox{and} 
~~~\tilde{\pi}^\beta_{cd}:=\frac{\partial 
\mathcal{L}}{\partial(\partial_t\tilde{\omega}^{cd}_\beta )}~,
\end{equation}
where $\mathcal{L}$ is the Lagrangian density of the action
(\ref{eq:prac}), and assuming the splitting conditions, one can then
show that
\begin{align}
\Pi^{\beta}_{cd}&=-4\alpha_0C^{t\beta}_{\:\:\:\:cd}~,
\\ \tilde{\pi}^\beta_{cd}&=2(\frac{1}{\kappa^2}-\frac{H^2}{12})e^t_{[c}e^\beta_{d]}~.
\end{align}
From the definition of the canonical momentum we get the constraints
$\Pi^t_{cd}=0$ and $\tilde{\pi}^t_{cd}=0$, which are primary first
class constraints and can be solved using the gauge fixing conditions
$\Omega_t^{\:ab}=0, ~ g^{ij}D_i \Omega_j^{\:ab}=0 $ and
$\tilde{\omega}_t^{\:ab}=0, ~ g^{ij}D_i
\tilde{\omega}_j^{\:ab}=0$. The remaining constraints
\begin{align}
\phi^t_0:=&p^t_0=0~,\\
\phi^i_I:=&p^i_I=0~, \label{eq:CC1}
\\ \phi_c:=&\Pi^i_{cd}e^d_i=0~,
\\ \varphi^j_c:=&\Pi^j_{cd}e^d_0-4\alpha_0C^{ij}_{\:\:cd}e^d_i=0~,\label{eq:CC3}\\ \phi^j_{cd}:=&\tilde{\pi}^j_{cd}-2(\frac{1}{\kappa^2}-\frac{H^2}{12})e^t_{[c}e^j_{d]}=0~.
\end{align}
are primary second class constraints, and are also obtained from the
definition of the canonical momentum. (We refer the reader to
Ref.~\cite{QGS} for more details on a constrained Hamiltonian system.)

In what follows, let $P, Q$ to stand for the canonical variables
and the symbol ``$\approx$" to denote the equality holding on the
surface spanned by all constraints, called the ``constraint surface'',
in short. Imposing all constraints, the Hamiltonian reads
\begin{align}
\mathcal{H}=&~P_I \partial_t Q^I-\mathcal{L}\nonumber\\ =&~\Pi^i_{cd}
\partial_t\Omega_i^{cd}+\tilde{\pi}^i_{cd}
\partial_t\tilde{\omega}_i^{cd}+p^\beta_c \partial_t
e^c_\beta+p_H\dot{H}+p_\psi\dot{\psi}-\mathcal{L}\nonumber\\ \approx
&~-\frac{1}{8\alpha_0}\Pi^i_{cd}\Pi_i^{cd}+\alpha_0
C^{ij}_{\:\:\:\:cd}C_{ij}^{\:\:\:\:cd}-\left(\frac{1}{\kappa^2}-\frac{H^2}{12}\right)R_{ij}^{\:\:\:\:cd}e^i_c
e^j_d+\mathcal{H}_{H,\psi}\nonumber\\ \approx &
~-\frac{1}{4\alpha_0}\Pi^i_{~0I}\Pi_i^{~0I}+\alpha_0
C^{ij}_{~~IJ}C_{ij}^{~~IJ}-\left[\frac{1}{8\alpha_0}\Pi^i_{IJ}\Pi_i^{IJ}-2\alpha_0
  C^{ij}_{~~0I}C_{ij}^{~~0I}\right]
\nonumber\\ &-(\frac{1}{\kappa^2}-\frac{H^2}{12})R_{ij}^{\:\:\:\:cd}e^i_c
e^j_d+\mathcal{H}_{H,\psi}\nonumber\\ \approx
&~-\frac{1}{4\alpha_0}\Pi^i_{~0I}\Pi_i^{~0I}+\alpha_0
C^{ij}_{~~IJ}C_{ij}^{~~IJ}
-\left(\frac{1}{\kappa^2}-\frac{H^2}{12}\right)R_{ij}^{\:\:\:\:cd}e^i_c
e^j_d+\mathcal{H}_{H,\psi}~,
\end{align}
where $p_H$ and $p_\psi$ are the canonical momenta of the scalar
field and the fermion field, respectively. Note that the term
$\left[\frac{1}{8\alpha_0}\Pi^i_{IJ}\Pi_i^{IJ}-2\alpha_0
  C^{ij}_{~~0I}C_{ij}^{~~0I}\right] $ vanish due to the symmetry
(\ref{eq:sym-2}) of the curvature tensor.

Denote the set of second class primary constraints by
$\Phi^A:=\{\phi^t_0,\phi^i_I,\phi_c,\varphi^j_c,\phi^j_{cd}\}$ and
define a new Hamiltonian density as
\begin{eqnarray}
\mathcal{H}_{\rm tot}:=\mathcal{H}+u_A\Phi^A,
\end{eqnarray}
where $u_A$ are Lagrange multipliers. All constraints need to satisfy
the consistency condition
\begin{equation}
0\approx \dot{\Phi}^A=\{\Phi^A,\mathbf{H}_{tot}\}~,
\end{equation} 
where $\mathbf{H}_{\rm tot}=\int{\mathcal{H}_{tot} e \textup{d}^3x}$ on
some equal time surface $\Sigma_t$. By imposing the consistency condition
on the constraints $\phi_c,\varphi^j_c,\phi^t_0$ and $\phi^i_I$ one
obtains the secondary constraint (the full details can be found in Appendix~C)
\begin{align}
\chi:=&~\frac{1}{2\alpha_0}\Pi^k_{0I}\Pi_{k}^{0I} + 2\alpha_0
C_{ij}^{\:\:\:\:lk}C^{ij}_{\:\:\:\:lk}+i\bar{\psi}\left(\gamma^I e^i_I
D_i \psi-2m\psi\right)-2\mu^2 H^2+2\lambda
H^4\nonumber\\ =&~0~.\label{eq:Chi}
\end{align}
Using the constraint (\ref{eq:Chi}) the Hamiltonian reads
\begin{align}
\mathcal{H}\approx &~2\alpha_0
C^{ijlk}C_{ijlk}-\frac{1}{2}i\bar{\psi}\gamma^I e^i_I D_i \psi
-(\frac{1}{\kappa^2}-\frac{H^2}{12})R_{ij}^{\:\:\:\:IJ}e^i_I
e^j_J+\frac{1}{2}g_{tt}p^2_H-\frac{1}{2}g^{ij}\partial_i H^\dagger
\partial_j H \nonumber\\ \approx
&~\mathcal{H}_{C^2}+\mathcal{H}_{\rm GR}~, \label{eq:Bounded}
\end{align}
where $\mathcal{H}_{C^2}$ and $\mathcal{H}_{\rm GR}$ are defined respectively as
\begin{align}
\mathcal{H}_{C^2}:=&2\alpha_0
C^{ijlk}C_{ijlk}-\frac{1}{2}i\bar{\psi}\gamma^I e^i_I D_i \psi~,
\\ \mathcal{H}_{\rm
  GR}:=&-(\frac{1}{\kappa^2}-\frac{H^2}{12})R_{ij}^{\:\:\:\:IJ}e^i_I
e^j_J+\frac{1}{2}g_{tt}p^2_H-\frac{1}{2}g^{ij}\partial_i H^\dagger
\partial_j H ~.
\end{align}
It is easy to show that $\mathcal{H}_{C^2}$ is bounded from below,
since the first term is positive definite and the second one is
proportional to the Hamiltonian of  a massless fermion.  To show that
$\mathcal{H}_{\rm GR}$ is also bounded from below, let us recall the
gauge fixing condition $~\omega_t^{~ab}=0~$, which implies
$~T_t^{~ab}=0$, since torsion is independent of the Levi-Civita spin
connection. Using Eq.~(\ref{eq:Riem-T}) we deduce that
\begin{align}
R_{ti}^{~~0I}=&~R_{ti}'^{~~0I}+\nabla_t T_i^{~0I}-\nabla_i T
_t^{~0I}+T_{t~J}^{~0}T_i^{~JI}-T_{i~J}^{~0}T_t^{~JI}\nonumber\\
=&~R_{ti}'^{~~0I}~,
\end{align}
while for $T_\mu^{~ab} \in \mathcal{T}_R$, the scalar curvature obtained by contracting Eq.~(\ref{eq:Riem-T}) reads
\begin{equation}
R=R'-||T||^2~.
\end{equation}
Hence, 
\begin{align}
R_{ij}^{~~IJ}e^i_I e^j_J =&~R-2R_{ti}^{~~0I}e^t_0
e^i_I\nonumber\\ =&~R'-||T||^2-2R_{ti}'^{~~0I}e^t_0 e^i_I
\nonumber\\ =&~(R'-2R_{ti}'^{~~0I}e^t_0 e^i_I)-(3
T_t^{~IJ}T^t_{~IJ}+T_{i~K}^{~I}T_j^{~KJ}e^i_I e^j_J)\nonumber\\ =&~
R'^{~~IJ}_{ij}e^i_I e^j_J-T_{i~K}^{~I}T_j^{~KJ}e^i_I
e^j_J~, \label{eq:Tor-GR}
\end{align}
which implies that $\mathcal{H}_{\rm GR}$ can be rewritten as
\begin{align}
\label{hgr}
\mathcal{H}_{\rm GR} \approx &
-(\frac{1}{\kappa^2}-\frac{H^2}{12})R^{'IJ}_{ij}e^i_I
e^j_J+\frac{1}{2}g_{tt}p^2_H-\frac{1}{2}g^{ij}\partial_i H^\dagger
\partial_j H
+(\frac{1}{\kappa^2}-\frac{H^2}{12})T_{i~K}^{~I}T_j^{KJ}e^i_I
e^j_J ~,
\end{align}
with a prime $'$ refering to torsion-free quantities.  Assuming the
Higgs field does not exceed the Planck mass, i.e. $H^2<12/\kappa^2$,
and noting that for the metric of signature $(+,-,-,-)$,
\begin{align}
T_{ijk}T^{ljk}=&g^{il}g^{jm}g^{kn}T_{ijk}T_{lmn} \nonumber\\ 
=&\sum^3_{i,j,k=1}g^{ii}g^{jj}g^{kk}T_{ijk}T_{ijk} \leq 0~,
\end{align}
one concludes that the last term in the {\sl r.h.s.} of
Eq.~(\ref{hgr}), which can be written as
$(\frac{1}{\kappa^2}-\frac{H^2}{12})T_{ijk}T^{ljk}$, is negative
definite, and therefore, unbounded from below. In contrast, the first
  two terms in the {\sl r.h.s.} of Eq.~(\ref{hgr}) are just the
  canonical Hamiltonian of the Palatini action in the presence of a
  scalar field interaction term~\cite{Pala-n}, leading to the
  classical dynamics of the Einstein-Hilbert action in the present of
  scalar field. We hence conclude that $\mathcal{H}_R$ is bounded
  from below if and only if torsion vanishes.

Finally, let us check whether the result agrees with Section~2. In the
vacuum case the constraint (\ref{eq:Chi}) becomes
\begin{equation}
\chi:=\frac{1}{2\alpha_0}\Pi^k_{0I}\Pi_{k}^{0I} + 2\alpha_0
C_{ij}^{\:\:\:\:lk}C^{ij}_{\:\:\:\:lk}=0~.\label{constr}
\end{equation}
Since $\Pi^k_{0I}\Pi_{k}^{0I}$ and
$C_{ij}^{\:\:\:\:lk}C^{ij}_{\:\:\:\:lk}$ are positive definite, the
constraint (\ref{constr}) implies that both terms have to vanish, and
hence the Hamiltonian reads
\begin{equation} 
\mathcal{H}\approx \mathcal{H}_{\rm GR}\approx
-\frac{1}{\kappa^2}R_{ij}^{\:\:\:\:cd}e^i_c e^j_d~.
\end{equation}
Hence, the Hamiltonian does not depend on the Weyl tensor,
in agreement with the fact that the vacuum case reduces to
Einstein gravity. Clearly then this Hamiltonian will give the same
dynamics as Einstein's equations in vacuum.

The above analysis can be easily applied in the spectral action. In
the simple vacuum case and considering a torsion field
$T_{\mu}^{~ab}\in \mathcal{T}_R$, the third order differential
equations can be reduced to the second order Einstein's
equations. Therefore, in this case the theory does not suffer from a
linear instability.  In the case of an almost commutative torsion
geometry and considering only matter fields whose Lagrangian do not
depend on the spin connection, one can still guarantee the stability of the
theory employing the method discussed in
Section~2. Moreover, if fermions and conformal invariant scalar fields
are present, the linear stability will still hold provided the
splitting conditions (\ref{eq:sp-1}) and (\ref{eq:sp-2}) are
satisfied. 
\section{Conclusions}

Noncommutative spectral geometry is a theoretical framework that can
offer a purely geometric explanation for the Standard Model of
particle physics.  The gravitational sector of the theory has terms
beyond the Einstein-Hilbert action and in particular it contains
higher derivative terms. Hence, one may wonder whether this
gravitational theory may be plagued by linear instabilities, namely
the appearance of negative energy modes.  We have
addressed this question here in two steps.

We have first considered the simple vacuum case and shown that
introducing a particular type of torsion, one can apply the method
presented in Ref.~\cite{Wheeler} and reduce the fourth order
differential equations in those of second order derived from vacuum
General Relativity, if and only if the torsion field vanishes.  We
have then cosnidered the spectral action of an almost commutative
torsion geometry.  For this latter case we have shown that one cannot
obtain the integrability condition in the presence of either fermion
fields or scalar fields. We have however argued that there exists a
class of almost commutative torsion geometry that leads to a
Hamiltonian which is bounded from below and hence argued that the theory does not
suffer from a local instability.

\appendix

\section{Spin connection}

The covariant derivative of a spinor $\psi$ or a tensor $V^a_\nu$ can be
expressed through the spin connection $\omega_\mu^{ab}$ as 
\begin{align}
D_\mu V^a_\nu=&\partial_\mu
V^a_\nu-\Gamma^\sigma_{\mu\nu}V^a_\sigma+\omega^{\:a}_{\mu\:\:b}V^b_\nu~,\\ D_\mu\psi=&\partial_\mu\psi+\frac{1}{4}\omega^{\:ab}_\mu\Sigma_{ab}\psi~,
\end{align}
respectively, where
$\Sigma_{ab}=\frac{1}{2}(\gamma_a\gamma_b-\gamma_b\gamma_a)$ and
$\Gamma^\sigma_{\mu\nu}$ stands for the affine connection defined as
\begin{equation}
\Gamma^\sigma_{\mu\nu}=e_{\nu,b}
e^\sigma_a\omega^{ab}_\mu-e_{\nu,b}\partial_\mu
e^{\sigma,b}~.\label{eq:G-Spin}
\end{equation}
The latter, Eq.~(\ref{eq:G-Spin}), can be rewritten as $D_\mu
e^a_\nu=0$, dubbed as the tetrad postulate. Note that the validity of
the tetrad postulate does not require to assume\\
\begin{itemize}
\item metric compatibility: $\omega^{ab}_\mu + \omega^{ba}_\mu=0$ 
\item torsion-free: $\Gamma^\sigma_{[\mu\nu]}=0$~. 
\end{itemize}
If the spin connection is metric compatible, then one can decompose
the curvature two-form into an irreducible representation of an
orthogonal group as follows
\begin{equation}
R_{\mu\nu}^{\:\:\:\:ab}=C_{\mu\nu}^{\:\:\:\:ab}+(e^{[a}_\mu
  R^{b]}_\nu-e^{[a}_\nu R^{b]}_\mu)-\frac{1}{3}Re^{[a}_\mu
  e^{b]}_\nu~,\label{eq:Weyl-Def}
\end{equation}
where $R^a_\mu=R_{\mu\alpha}^{\:\:\:\:ab}e^\alpha_b $ and $R=
R_{\mu\nu}^{\:\:\:\:ab}e^\mu_a e^\nu_b$. Hence, in the coordinate basis, one has
\begin{align}
R_{\mu\nu\rho\sigma}=&R_{\mu\nu}^{\:\:\:\:ab}e_{a,\rho}
e_{b,\sigma}\nonumber\\ =&C_{\mu\nu}^{\:\:\:\:ab}e_{a,\rho}
e_{b,\sigma}+(e^{[a}_\mu R^{b]}_\nu-e^{[a}_\nu R^{b]}_\mu)e_{a,\rho}
e_{b,\sigma}-\frac{1}{3}Re^{[a}_\mu e^{b]}_\nu e_{a,\rho}
e_{b,\sigma}\nonumber\\ =&C_{\mu\nu\rho\sigma}
+(g_{\mu[\rho}R_{\sigma]\nu}-g_{\nu[\rho}R_{\sigma]\mu})-\frac{1}{3}Rg_{\mu[\rho}g_{\sigma]\nu}~.
\end{align}
Assuming also that the spin connection is torsion-free, one concludes
that $C_{\mu\nu\rho\sigma},\ R_{\mu\nu}$ and $R$ become the Weyl
tensor, the Ricci tensor and the Ricci scalar, respectively.  
\section{Equivalent linearized actions}

We will show that the linearized theories obtained from (i) the action
(\ref{eq:NCG Kibble}) and (ii) the spectral action with
torsion~\cite{gr-TSA}, are equivalent. First let us write down the
spectral action with torsion, which we will denote by $S_{\rm TS}$.
For a torsion tensor $T_{\mu\nu\sigma} \in \mathcal{T}_R$, we have by
definition that
\begin{equation}
0=R_{\mu\nu\rho\sigma}-R_{\rho\sigma\mu\nu}=\frac{1}{2}(dT)_{\mu\nu\rho\sigma}-\nabla_\rho
T_{\sigma\mu\nu}+\nabla_\sigma T_{\rho\mu\nu}~, \label{eq:anti-Riem}
\end{equation}
and
\begin{equation}
0=R_{\mu\rho}-R_{\rho\mu}=g^{\nu\sigma}\left(R_{\mu\nu\rho\sigma}-R_{\rho\sigma\mu\nu}\right)=\nabla_\sigma
T^\sigma_{~\rho\mu}~. \label{eq:anti-Ric}
\end{equation}
Hence, the spectral action (modulo the Euler characteristic number) is
reduced to
\begin{align}
S_{\rm TS}\sim & f_4 \Lambda^4 a_0(\mathcal{D}^2)+f_2 \Lambda^2
a_2(\mathcal{D}^2) + f(0) a_4(\mathcal{D}^2)\nonumber\\ \sim & \int
\sqrt{|g|}d^4x\left( \alpha_2 \Lambda^4 +
\frac{1}{\kappa^2}(R'-||T||^2) -\alpha_0||C'||^2\right)~.
\end{align}
Note that the torsion tensor
$T_{\mu\nu\sigma}:=3\tilde{T}_{\mu\nu\sigma}$, where
$\tilde{T}_{\mu\nu\sigma}$ denotes the torsion defined in
Ref.~\cite{gr-TSA}. To compare $S_{\rm gr}$ with $S_{\rm TS}$, we will
write explicitly the torsion terms which are contained in $S_{\rm
  gr}$. Consider the square of the traceless tensor
$C_{\mu\nu}^{\:\:\:\:ab}$ defined in Eq.~(\ref{eq:Weyl-Def}):
\begin{align}
||C||^2=&||R_{\mu\nu\rho\sigma}||^2-2||R_{\mu\nu}||+\frac{1}{3}R^2 \nonumber\\
         =&||R'_{\mu\nu\rho\sigma}||^2+\frac{1}{4}||dT||^2-\frac{1}{3}R'||T||^2+4B(T)+\frac{1}{3}||T||^4 \nonumber\\
          &-2\left(||R'_{\mu\nu}||+\frac{1}{3}||T||^4-\frac{1}{2}R'||T||^2+2B(T)\right)\nonumber\\
          &+\frac{1}{3}\left(R'^2-2R'||T||^2+||T||^4\right)\nonumber\\
         =&||C'||^2+\frac{1}{4}||dT||^2~,
\end{align}
where $B(T):=-R'_{\mu\nu}T^{\mu\sigma\rho}
T^\nu_{~\sigma\rho}+\frac{1}{4}R'||T||^2$ and the curvature scalar $R$
is $R=R'-||T||^2.$ Substituting $||C||^2$ and $R$ in the action
(\ref{eq:NCG Kibble}) we get
\begin{align}
S_{\rm gr}= \int  \sqrt{|g|}\left[\alpha_2\Lambda
 ^4+\frac{1}{\kappa^2} \left(R'-||T||^2\right)-\alpha_0
 \left(||C'||^2+\frac{1}{4}||dT||^2\right)\right]d^4x~.
 \end{align}
Using Eq.~(\ref{eq:anti-Riem}) we rewrite $||dT||^2$ as
 \begin{align}
 ||dT||^2=&~(dT)_{\mu\nu\rho\sigma}(dT)^{\rho\sigma\mu\nu}\nonumber\\ =&~(-\nabla_\rho
 T_{\sigma\mu\nu}+\nabla_\sigma T_{\rho\mu\nu})( -\nabla^\mu
 T^{\nu\rho\sigma}+\nabla^\nu
 T^{\mu\rho\sigma})\nonumber\\ =&~4\nabla_\rho
 T_{\sigma\mu\nu}\nabla^\mu
 T^{\nu\rho\sigma}\nonumber\\ =&~4\nabla^\mu
 (T^{\nu\rho\sigma}\nabla_\rho
 T_{\sigma\mu\nu})-4T^{\nu\rho\sigma}\nabla^\mu\nabla_\rho
 T_{\sigma\mu\nu}\nonumber\\ =&~4\nabla^\mu
 (T^{\nu\rho\sigma}\nabla_\rho
 T_{\sigma\mu\nu})+4T^{\nu\rho\sigma}\nabla_\rho \nabla^\mu
 T_{\sigma\mu\nu}-4T^{\nu\rho\sigma}[\nabla^\mu,\nabla_\rho]
 T_{\sigma\mu\nu}\nonumber\\ =&4\nabla^\mu
 (T^{\nu\rho\sigma}\nabla_\rho
 T_{\sigma\mu\nu})-8T^{\nu\rho\sigma}\left(R'^{\mu}_{~\rho\sigma\alpha}-\frac{1}{2}\delta^\mu_\sigma
 R'_{\rho\alpha}\right)T^{\alpha}_{~\mu\nu}~.
 \end{align}
Note that to obtain the last line we have used that the divergent of a
torsion field vanishes (Eq.~(\ref{eq:anti-Ric})) and the identity
 \begin{equation}
 [\nabla_\mu,\nabla_\nu]V_{\rho\sigma\alpha} = R'^{~~~\beta}_{\mu\nu\rho}V_{\beta\sigma\alpha}+ R'^{~~~\beta}_{\mu\nu\sigma}V_{\rho\beta\alpha}+ R'^{~~~\beta}_{\mu\nu\alpha}V_{\rho\sigma\beta}~.
 \end{equation}
Thus, the action $S_{\rm gr}$ reads
 \begin{eqnarray}
\label{action-terms}
S_{\rm gr}&=&\int \sqrt{|g|}\left[\alpha_2\Lambda
  ^4+\frac{1}{\kappa^2} \left(R'-||T||^2\right)- \alpha_0
  ||C'||^2\right]d^4x + 2\int \sqrt{|g|}
T^{\nu\rho\sigma}\left(R'^{\mu}_{~\rho\sigma\alpha}-\frac{1}{2}\delta^\mu_\sigma
R'_{\rho\alpha}\right)T^{\alpha}_{~\mu\nu} d^4x\nonumber\\ &=&S_{\rm ST}+ 2\int \sqrt{|g|}
T^{\nu\rho\sigma}\left(R'^{\mu}_{~\rho\sigma\alpha}-\frac{1}{2}\delta^\mu_\sigma
R'_{\rho\alpha}\right)T^{\alpha}_{~\mu\nu} d^4x~.
 \end{eqnarray}
Since the terms in the integrand appearing in the {\sl
  r.h.s.} of Eq.~(\ref{action-terms}) are of order ${\cal
  O}(\omega^3)$, they can be discarded in the linearized theory. Thus,
the actions $S_{\rm gr}$ and $S_{\rm TS}$ lead to theories which are
equivalent in linear order.  \\\\
\section{Solving the constraints}

For a constrained Hamiltonian system, the time evolution of any phase
space function $f(P(x), Q(x))$ is defined by the Poisson bracket of
$f$ with the Hamiltonian:
\begin{align}
\{f(x),\mathbf{H}_{\rm tot}\}=&\int{d^3y}\{f(x), e\mathcal{H}_{\rm
  tot}(y)\}_{x_0=y_0} \nonumber\\ &\int{d^3yd^3z}\left(\frac{\partial
  f(x)}{\partial Q(z)}\frac{\partial e\mathcal{H}_{tot}(y)}{\partial
  P(z)}-\frac{\partial f(x)}{\partial P(z)}\frac{\partial e
  \mathcal{H}{\rm tot}(y)}{\partial Q(z)}\right)_{x_0=y_0}~.
\end{align}
The consistency condition requires that the constraints do not have a
time evolution on the constraint surface.

At this point let us make a remark that will be useful later. Denoting
by $\Phi^A$ the set of second class primary constraints, one has 
\begin{align}
 0=&~\dot{\Phi}^A=\{\Phi^A, e \mathcal{H}_{\rm tot}\}
=\{\Phi^A, e \mathcal{H} + e u_B\Phi^B\}\nonumber\\
=&~\{\Phi^A,
 e\mathcal{H}\}+e
 u_B\{\Phi^A,\Phi^B\}+u_B\{\Phi^A,e\}\Phi^B\nonumber\\ \approx &~
 e\left(\frac{1}{e}\{\Phi^A,
 e\mathcal{H}\}+u_B\{\Phi^A,\Phi^B\}\right)~,\label{eq:Cons}
\end{align}
where $u_B$ stand for Lagrange multipliers.  Hence, if the quantity
$\left(\frac{1}{e}\{\Phi^A,
e\mathcal{H}\}+u_B\{\Phi^A,\Phi^B\}\right)$ is weakly equal to zero,
then the consistency condition is satisfied. From Eq.~(\ref{eq:Cons})
one may either obtain the Lagrange multiplier $u_B$ or a new
constraint, which is not a linear combination of the primary
constraints. This new constraint is called the {\sl secondary
  constraint} and we will define it by $\chi=0$.

In what follows we will derive the constraint (\ref{eq:Chi}). 
Note that we use the identities
\begin{align}
\delta e^\mu_a =& -e^\mu_b e^\nu_a\delta e^b_\nu~, \\
\delta e =& ~e e^\mu_a \delta e^a_\mu = -e e^a_\mu\delta e^\mu_a~.
\end{align} 
Let us first reduce the number of unknown Lagrange multipliers by
imposing the consistency condition on the constraints $\phi_c=0$ and
$\varphi^j_c=0$.  \\\\ $\bullet$
~$0=\dot{\phi}_c=\{\phi_c,e\mathcal{H}_{\rm tot}\}:$ \\ Using
Eq.~(\ref{eq:Cons}) the consistency condition implies
\begin{align}
0\approx~\{\phi_c,\mathcal{H}\}+u^0_t \{\phi_c,\phi^t_0\}+u^I_i
\{\phi_c,\phi^i_I\}+w^a_j\{\phi_c,\varphi^j_a\}~.
\end{align}
Contraction with $e^c_t=(e^0_t,0,0,0)$ then yields
\begin{align}
0\approx &~\{e^c_t\phi_c,\mathcal{H}\}+u^0_t
e^c_t\{\phi_c,\phi^t_0\}+u^I_i e^c_t\{\phi_c,\phi^i_I\}+w^a_j
e^c_t\{\phi_c,\varphi^j_a\}
\nonumber\\ \approx&~\{e^0_t\phi_0,\mathcal{H}\}+u^0_t
\{e^0_t\phi_0,\phi^t_0\}-u^0_t \phi_0\{e^0_t,\phi^t_0\} +u^I_i
\{e^0_t\phi_0,\phi^i_I\}+w^j_a e^0_t\{\phi_0,\varphi^a_j\}
\nonumber\\ \approx&~\{e^0_t\phi_0,\mathcal{H}\}+u^0_t
\{e^0_t\phi_0,\phi^t_0\}+u^I_i \{e^0_t\phi_0,\phi^i_I\}+w^a_j
e^0_t\{\phi_0,\varphi^j_a\} ~. \label{eq:C1}
\end{align}
$\bullet$
~ $0=\dot\varphi^j_J=\{\varphi^j_J,e\mathcal{H}_{tot}\}:$
\begin{align}
0\approx&~\{\varphi^j_J,\mathcal{H}\}+u^0_t\{\varphi^j_J,\phi^t_0\}+u^I_i
\{\varphi^j_J,\phi^i_I\}-u_c\{\phi_c,\varphi^j_J\} ~.
\end{align}
Contraction with $e^J_j$ then yields
\begin{equation}
0\approx
~\{e^J_j\varphi^j_J,\mathcal{H}\}+u^0_t\{e^J_j\varphi^j_J,\phi^t_0\}+u^I_i
\{e^J_j\varphi^j_J,\phi^i_I\}-u_c e^J_j\{\phi_c,\varphi^j_J\}
~. \label{eq:C2}
\end{equation}
Combining Eqs.~(\ref{eq:C1}),(\ref{eq:C2}) and using $~e^J_j
\varphi^j_J= e^0_t\phi_0~$, one gets
\begin{equation}
u^ce^J_j=-w^J_j e^c_t~.
\end{equation}
Defining the scalar $C:=\frac{1}{3}w^J_j e^j_J$ one then obtains
\begin{equation}
u^a=-Ce^a_t~, \:\:\:\:\:\: w^J_j=Ce^J_j~. \label{eq:Lm}
\end{equation}
As a consequence of (\ref{eq:Lm}) the total Hamiltonian is reduced to
\begin{align}
\mathcal{H}_{tot}=&~\mathcal{H}-Ce^a_t\phi_a+Ce^J_j\varphi^j_J+u^0_t\phi^t_0+u^I_i\phi^i_I+u^{ab}_j
\phi^j_{ab}\nonumber\\ =&\mathcal{H}+u^0_t\phi^t_0+u^I_i\phi^i_I+u^{ab}_j
\phi^j_{ab}.
\end{align}
Next, to obtain the constraint Eq.~(\ref{eq:Chi}), we analyze the
consistency of the constraints $\phi^t_0=0$ and $\phi^i_I=0$.
\\\\ $\bullet$ ~ $0=\dot{\phi}^t_0=\{\phi^t_0,e\mathcal{H}_{\rm
  tot}\}:$\\ We have
\begin{align}
0=&\frac{1}{e}\{\phi^t_0,e\mathcal{H}\}+u^a\{p^t_0,\phi_a\}+w^a_j\{p^t_0,\varphi^j_a\}+u^{ab}_j\{p^t_0,\phi^j_{ab}\}\nonumber\\
 \approx &~ \{p^t_0,\mathcal{H}\}+\frac{1}{e}\mathcal{H}\{p^t_0,e\}-2u^{0J}_j\left(\frac{1}{\kappa^2}-\frac{H^2}{12}\right)(e^t_0)^2e^j_J\nonumber\\
 \approx&~ \frac{1}{2\alpha_0} e^t_0 \Pi^i_{0K}\Pi_i^{0K}+\{p^t_0, \mathcal{H}_{H,\psi}\}-e^t_0\mathcal{H}-2u^{0J}_j\left(\frac{1}{\kappa^2}-\frac{H^2}{12}\right)(e^t_0)^2e^j_J \nonumber \\
 \approx&~ \frac{3}{4\alpha_0} e^t_0 \Pi^i_{0K}\Pi_i^{0K}-\alpha_0 e^t_0 C_{ij}^{~~IJ}C^{ij}_{~~IJ}+\left(\frac{1}{\kappa^2}-\frac{H^2}{12}\right)e^t_0 R_{ij}^{IJ}e^i_I e^j_J\nonumber\\
 &+\left( \{p^t_0, \mathcal{H}_{H,\psi}\}-e^t_0\mathcal{H}_{H.\psi}\right)-2u^{0J}_j\left(\frac{1}{\kappa^2}-\frac{H^2}{12}\right)(e^t_0)^2e^j_J ~.  \label{eq:EOM1}
\end{align}
Multiplying the above equation, Eq.~(\ref{eq:EOM1}), with $e^0_t$ we obtain 
\begin{align}
 \approx&~ \frac{3}{4\alpha_0} \Pi^i_{0K}\Pi_i^{0K}-\alpha_0
 C_{ij}^{~~IJ}C^{ij}_{~~IJ}+\left(\frac{1}{\kappa^2}-\frac{H^2}{12}\right)R_{ij}^{IJ}e^i_I
 e^j_J\nonumber\\ &+\left( e^0_t\{p^t_0,
 \mathcal{H}_{H,\psi}\}-\mathcal{H}_{H.\psi}\right)-2u^{0J}_j\left(\frac{1}{\kappa^2}-\frac{H^2}{12}\right)e^t_0
 e^j_J ~.\label{eq:u}
\end{align}
\\\\
$\bullet$~ 
$0=\dot{\phi}^k_K=\{\phi^k_K,e\mathcal{H}_{tot}\}:$\\
We have
\begin{align}
0=&\frac{1}{e}\{p^k_K,e\mathcal{H}\}+u^a\{p^k_K,\phi_a\}+w^a_j\{p^k_K,\varphi^j_a\}+u^{ab}_j\{p^k_K,\phi^j_{ab}\}\nonumber\\ \approx
&~
\{p^k_K,\mathcal{H}\}+\frac{1}{e}\mathcal{H}\{p^k_K,e\}-2u^{0J}_j\left(\frac{1}{\kappa^2}-\frac{H^2}{12}\right)e^t_0
e^k_J
e^j_K\nonumber\\ \approx&~\frac{1}{2\alpha_0}\Pi^k_{0I}\Pi_{j}^{0I}e^j_K+4\alpha_0
e^m_K
C^{kl}_{~~IJ}C_{ml}^{~~IJ}-2\left(\frac{1}{\kappa^2}-\frac{H^2}{12}\right)R_{ij}^{~~IJ}e^i_I
e^k_J e^j_K\nonumber \\ &
+\{p^k_K,\mathcal{H}_{H,\psi}\}-e^k_K\mathcal{H}-2u^{0J}_j\left(\frac{1}{\kappa^2}-\frac{H^2}{12}\right)e^t_0
e^k_J e^j_K \nonumber\\ \approx
&~\frac{1}{2\alpha_0}\Pi^k_{0I}\Pi_{j}^{0I}e^j_K+\frac{1}{4\alpha_0}e^k_K\Pi^i_{0I}\Pi_{i}^{0I}+4\alpha_0\left(
e^m_K C^{kl}_{~~IJ}C_{ml}^{~~IJ}-\frac{1}{4}e^k_K
C^{ij}_{~~IJ}C_{ij}^{~~IJ}\right)\nonumber\\ &-2\left(\frac{1}{\kappa^2}-\frac{H^2}{12}\right)\left(R_{ij}^{~~IJ}e^i_I
e^k_J e^j_K-\frac{1}{2}e^k_K R_{ij}^{~~IJ}e^i_I e^j_J
\right)+\{p^k_K,\mathcal{H}_{H,\psi}\}-e^k_K\mathcal{H}_{H,\psi}\nonumber\\ &-2u^{0J}_j\left(\frac{1}{\kappa^2}-\frac{H^2}{12}\right)e^t_0
e^k_J e^j_K ~.
\end{align}
Contracting with $e^K_k$ we obtain 
\begin{align}
 0 \approx&~\frac{5}{4\alpha_0}\Pi^k_{0I}\Pi_{k}^{0I} +\alpha_0 C_{ij}^{~~IJ}C^{ij}_{~~IJ}+\left(\frac{1}{\kappa^2}-\frac{H^2}{12}\right)R_{ij}^{~~IJ}e^i_I e^j_J\nonumber\\
 &+e^K_k\{p^k_K,\mathcal{H}_{H,\psi}\}-3\mathcal{H}_{H,\psi}-2u^{0J}_j\left(\frac{1}{\kappa^2}-\frac{H^2}{12}\right)e^t_0 e^j_J~.\label{eq:2-nd CC}
\end{align}
Combining Eqs.~(\ref{eq:2-nd CC}) and Eq.~(\ref{eq:u}) we have a constraint equation
\begin{align}
0\approx&~\frac{1}{2\alpha_0}\Pi^k_{0I}\Pi_{k}^{0I} + 2\alpha_0
C_{ij}^{\:\:\:\:lk}C^{ij}_{\:\:\:\:lk}+\frac{1}{4\alpha_0}\Pi^k_{IJ}\Pi_{k}^{IJ}+
4\alpha_0 C_{ij}^{~~0I}C^{ij}_{~~0I}
\nonumber\\ &~+e^K_k\{p^k_K,\mathcal{H}_{H,\psi}\}- e^0_t\{p^t_0,
\mathcal{H}_{H,\psi}\}-2\mathcal{H}_{H,\psi}
\nonumber\\ \approx&~\frac{1}{2\alpha_0}\Pi^k_{0I}\Pi_{k}^{0I} + 2\alpha_0
C_{ij}^{\:\:\:\:lk}C^{ij}_{\:\:\:\:lk}+i\bar{\psi}\left(\gamma^I e^i_I
D_i \psi-2m\psi\right)-2\mu^2 H^2+2\lambda H^4\nonumber\\
=&:~\chi~,  \label{eq:2-Coco}
\end{align}
which is not a linear combination of the primary constraints.  In
conclusion, $\chi=0$ is a secondary constraint, which arises from the
consistency condition.


\end{document}